\documentclass{article}

\usepackage{arxiv}

\usepackage[utf8]{inputenc} 
\usepackage[T1]{fontenc}    
\usepackage{hyperref}       
\usepackage{url}            
\usepackage{booktabs}       
\usepackage{amsfonts}       
\usepackage{nicefrac}       
\usepackage{microtype}      
\usepackage{lipsum}		
\usepackage{graphicx}
\usepackage{amsmath}
\usepackage{amssymb}
\usepackage{doi}

\pdfoutput=1

\title{Long-term future particle accelerators}

\author{ \href{https://orcid.org/0000-0002-1178-5136}{\includegraphics[scale=0.06]{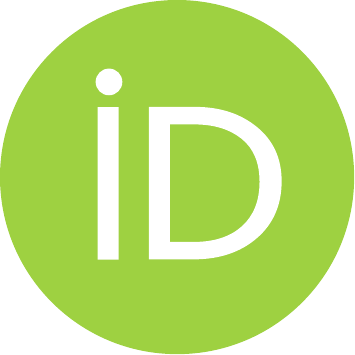}\hspace{1mm}Javier Resta L\'opez}\thanks{The author acknowledges support by the Generalitat Valenciana under Grant agreement No. CIDEGENT/2019/058.} \\
	Institute of Materials Science (ICMUV)\\
	University of Valencia\\
	46071, Paterna, Spain \\
	\texttt{javier2.resta@uv.es} \\
}

\hypersetup{
pdftitle={A template for the arxiv style},
pdfsubject={q-bio.NC, q-bio.QM},
pdfauthor={David S.~Hippocampus, Elias D.~Striatum},
pdfkeywords={First keyword, Second keyword, More},
}

\begin{document}
\maketitle

\begin{abstract}
Particle accelerators have enabled forefront research in high energy physics and other research areas for more than half a century. Accelerators have directly contributed to 26 Nobel Prizes in Physics since 1939 as well as another 20 Nobel Prizes in Chemistry, Medicine and Physics with X-rays. Although high energy physics has been the main driving force for the development of the particle accelerators, accelerator facilities have continually been expanding applications in many areas of research and technology. For instance, active areas of accelerator applications include radiotherapy to treat cancer, production of short-lived medical isotopes, synchrotron light sources, free-electron lasers, beam lithography for microcircuits, thin-film technology and radiation processing of food. Currently, the largest and most powerful accelerator is the Large Hadron Collider (LHC) at CERN, which accelerates protons to multi-TeV energies in a 27 km high-vacuum ring. To go beyond the maximum capabilities of the LHC, the next generation of circular and linear particle colliders under consideration, based on radiofrequency acceleration, will require multi-billion investment, kilometric infrastructure and a massive power consumption. These factors pose serious challenges in an increasingly resource-limited world. Therefore, it is important to look for alternative and sustainable acceleration techniques. This review article pays special attention to novel accelerator techniques to overcome present acceleration limitations towards more compact and cost-effective long term future accelerators.
\end{abstract}

\keywords{High energy collider \and Plasma wakefield \and Dielectric accelerator \and Solid-state plasma \and Plasmonic accelerator}

\section{Introduction} 

Currently, there are more than 30,000 particle accelerators in operation around the world, with a strong impact in science, industry and economy. Although historically high energy physics (HEP) has been the main driving force for the accelerator R$\&$D, only $1\%$ of particle accelerators worldwide are used for research purposes, the rest are mostly used to support commercial, industrial and medical work.  For instance, active areas of accelerator applications include radiotherapy to treat cancer, production of short-lived medical isotopes, synchrotron light sources, free-electron lasers, beam lithography for microcircuits, thin-film technology and radiation processing of food.

Since the pioneering work by Gustave Ising, Rolf Wider\o{}e, Ernest Lawrence and many others in the 1920s and 1930s, the accelerator technology has progressed immensely \cite{History}. Particularly during the past 70 years, an impressive development of high energy colliders, with ever increasing luminosities and centre-of-mass energies, has resulted in a number of fundamental discoveries in particle physics. In recent decades, experiments at hadron colliders such as the Super Proton Synchrotron (SPS) and Tevatron, at $e^{+}e^{-}$ colliders such as the Large Electron-Positron Collider (LEP) and the Stanford Linear Collider (SLC), and at the electron-proton collider HERA (Hadron-Electron Ring Accelerator) have explored the energy range up to several 100~GeV and established beyond doubt the validity of the Standard Model in this range of energies \cite{FZ-VS}. In this tremendous progress the development and optimisation of the radiofrequency (RF) technology has played a key role. Nowadays most particle accelerators are driven by RF electromagnetic fields. Currently, the largest and most powerful RF-driven accelerator in operation is the Large Hadron Collider (LHC) at CERN \cite{LHCreport}, which accelerates protons to multi-TeV energies in a 27 km high-vacuum ring. Its kilometric size and complexity have earned it the name of the "Cathedral" of the particle physics. With the discovery of the Higgs boson in 2012, the LHC has contributed to complete the Standard Model \cite{Higgs1, Higgs2}.

To go beyond the maximum energy capabilities of the LHC, the next generation of RF-driven colliders under consideration would be among the largest and most complex facilities built on Earth, requiring intensive R$\&$D and multi-billion Euro investments. For instance, the long-term goal of the Future Circular Collider (FCC) study hosted by CERN \cite{FCC1, FCC2} is the design and construction of a 100 TeV hadron collider in a 100 km long tunnel. Furthermore, future linear $e^{+}e^{-}$ colliders based on RF technology, such as the International Linear Collider (ILC) \cite{ILC} or the Compact Linear Collider (CLIC) \cite{CLIC}, are designed to produce acceleration gradients of between 30 MV/m (ILC) and 150 MV/m (CLIC, 30 GHz frequency operation mode). These machines must therefore be tens of kilometres long to reach the desired beam energies, 125 GeV (ILC) and 1.5 TeV (CLIC). Other big project proposed by the HEP community is a high luminosity multi-TeV muon collider \cite{MC}. However, practically all the projects mentioned above, based on conventional RF acceleration technology, present serious challenges in terms of size and cost per GeV of beam energy. They need large-scale infrastructure (10--100 km scale) and, in most of the cases, a billion investment. To afford such a cost, it usually requires a consortium of several countries. Another key challenge is the sustainability in terms of the AC wall plug power consumption.  High energy colliders are really "hungry energy" machines that require hundreds of MW AC wall power to operate.  To reduce such a power consumption in the latest years there have been a trend towards more "green accelerators".  This concept includes the investigation of different ways to reduce energy consumption. For instance, through the improvement of the efficiency of existing technologies, e.g. increasing the efficiency of klystrons \cite{Cai:2019}. Also the design of Energy-Recovery Linacs (ERLs) can be considered as part of the R$\&$D effort aimed  at increasing energy efficiency \cite{Litvinenko:2020}. 

To tackle the limitations of conventional accelerators, nowadays there is also an intensive  R$\&$D program on novel advanced concepts, some of which we briefly describe in the next sections. These alternative acceleration techniques might define the long term future of particle accelerators, thus transforming the current paradigm in collider development towards more sustainable, compact and low-cost machines. For a more exhaustive review of past, present and future accelerators and colliders, see for example Refs.~\cite{VS, FZ-VS, Ferrario}.

\section{Plasma wakefield}

Since 1930s the RF acceleration technology has practically dictated the development of high energy colliders. Their power, cost and size have evolved with the continuous improvements and optimisation experienced by RF cavities. However, over the recent decades it has become apparent that RF technology is reaching its limits in terms of achievable accelerating gradients. Currently the maximum electromagnetic fields that it can support are limited to approximately 100 MV/m due to surface breakdown \cite{Abe:2016}. To overcome this limit, Tajima and Dawson \cite{Tajima:1979} proposed an alternative solution, a plasma accelerator based on laser-driven wakefields. Since a plasma\footnote{A plasma is defined as a fluid of positive and negative charges.} is already a broken-down medium, there is no breakdown limit, compared to the conventional metallic RF cavities.  

\begin{figure}[!htb]\centering
	\includegraphics[width=12cm]{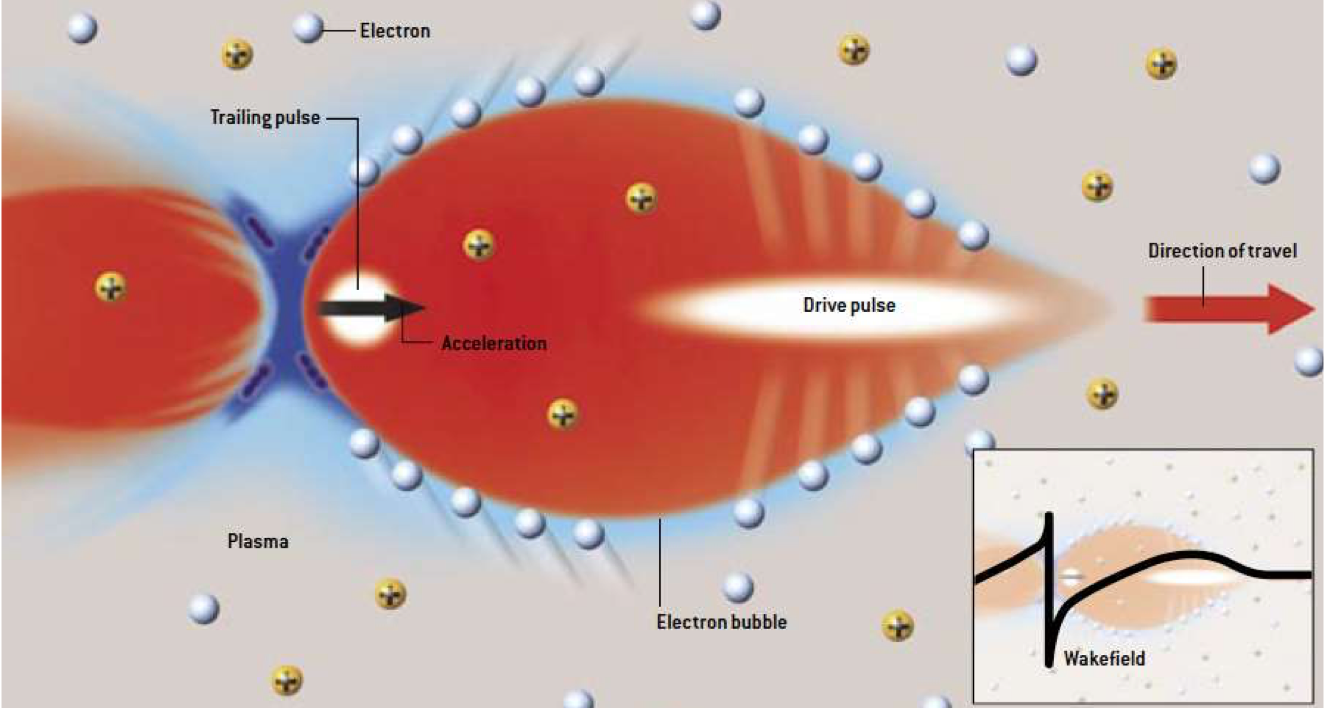}
	\caption{Conceptual drawing of a plasma wakefield accelerator \cite{Joshi:2006}.}
\label{fig:plasmacc}
\end{figure}

In general terms, the concept of plasma wakefield acceleration is illustrated in Figure~\ref{fig:plasmacc}. A drive pulse enters the plasma and expels the electrons of the plasma outward. The plasma ions move a negligible amount due to their higher mass with respect to the electrons. In consequence, a positively charged ion channel is formed along the path of the drive pulse. The drive pulse can be either a short laser pulse (Laser Wakefield Acceleration (LWFA)) \cite{Tajima:1979} or an electron or proton bunch (Plasma Wakefield Acceleration (PWFA)) \cite{Chen:1985}. After the passage of the driver (laser or particle beam), the plasma electrons rush back in, attracted by the transverse restoring force of the ion channel. In this way a space charge driven oscillation is excited, generating alternating regions of negative and positive charge, thus inducing a strong longitudinal electric field behind the driver, the so-called plasma wakefield. Therefore, if a witness charged particle bunch is injected behind the driver at the correct distance and phase, then it will be accelerated with high gradients, in some cases exceeding 100 GV/m.

In the last decades, acceleration of relativistic electrons in gaseous plasma has been successfully demonstrated in several experiments, see for example \cite{Gordon:1998, Leemans:2014, Litos:2015}. In the same way, plasma wakefield acceleration of positrons has also been demonstrated \cite{Blue:2003, Corde:2015, Doche:2017}. Furthermore, the Advanced Wakefield Experiment (AWAKE) at CERN has recently shown that 10--20~MeV electrons can be accelerated to GeV energies in a plasma wakefield driven by a highly relativistic self-modulated proton bunch \cite{AWAKE}.

Previously we have mentioned injection of a witness beam to be accelerated by the induced wakefield. In addition, internal injection of electrons from the plasma itself is also possible in the so-called "bubble regime" \cite{Geddes:2004, Mangles:2004, Faure:2004}.

The wakefields generated in plasma can be evaluated by analytical expressions \cite{Esarey:2009}. The maximum accelerating field in a plasma accelerator can be estimated as follows:

\begin{equation}
\label{eq:longwake}
E_z  [\textrm{V/m}]\approx 96 \sqrt{n_e [\textrm{cm}^{-3}]}
\end{equation}

\noindent where $n_e$ is the electron density in the plasma. This density determines the plasma frequency, $\omega_p=\sqrt{e^2n_e/m_e \varepsilon_0}$, with $m_e$ and $e$ the electron rest mass and charge, respectively, and $\varepsilon_0$ the electrical permittivity of free space. For example, a typical plasma density $n_e=10^{18}$~cm$^{-3}$ gives $E_z \sim 100$~GV/m, i.e. approximately three orders of magnitude higher that the maximum gradient obtained in RF structures. 

\begin{figure}[!htb]\centering
	\includegraphics[width=12cm]{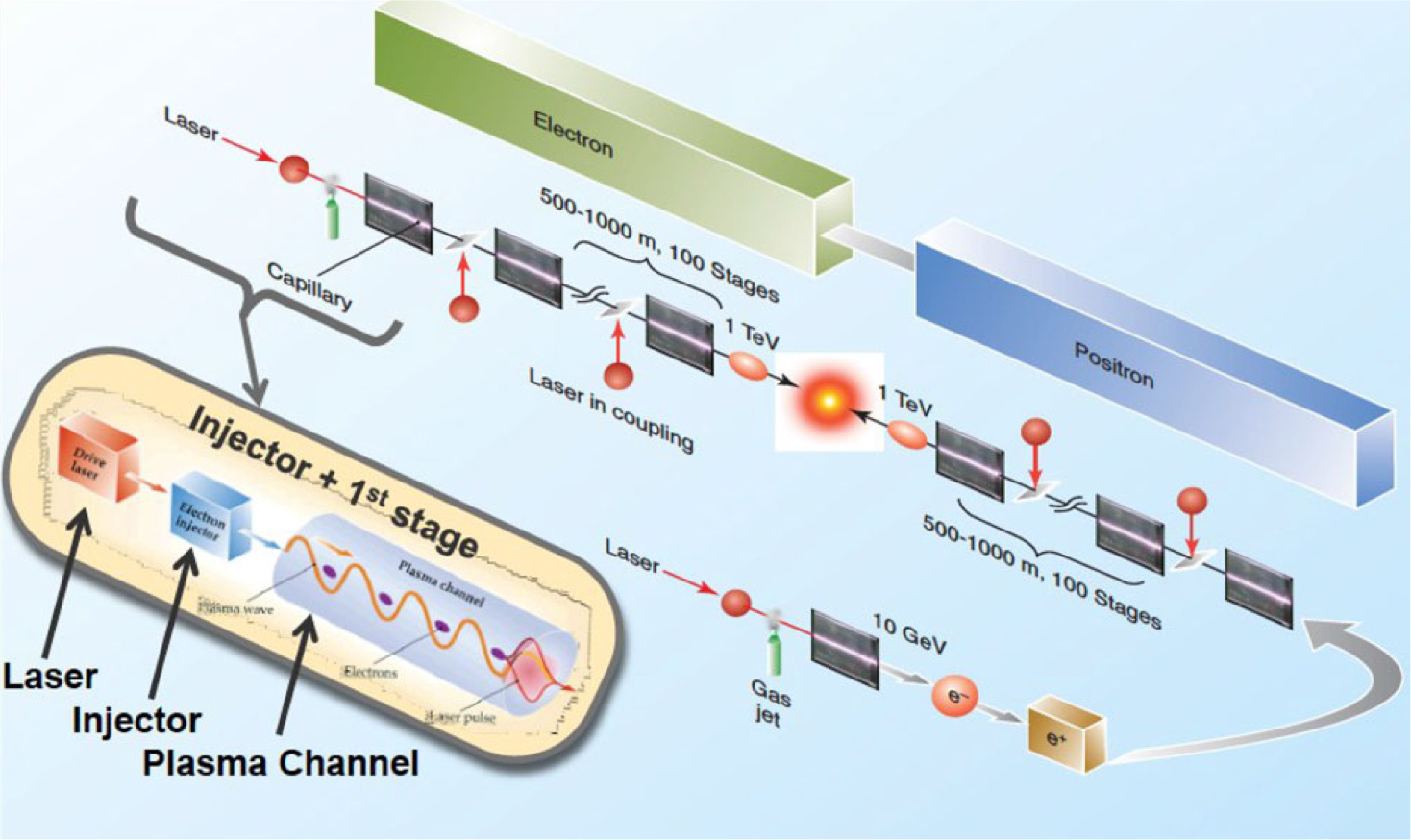}
	\caption{Conceptual design of a LWFA based linear collider \cite{Schroeder:2010}.}
\label{fig:wakecollider}
\end{figure}

Multi-TeV $e^{+}e^{-}$ concepts based on staged plasma cells have been proposed \cite{Schroeder:2010}, see Figure~\ref{fig:wakecollider}. However, one of the main drawbacks of plasma accelerators is the production of relatively poorer beam quality (higher energy spread and transverse size) with respect to conventional RF techniques. In order to tackle this problem and produce a reliable and competitive plasma-based accelerator technology, a big research consortium, EuPRAXIA, was recently established \cite{Assmann:2019}. A comprehensive review of near and long-term potential applications of plasma accelerators can be found in \cite{Joshi:2020}.

\section{Solid-state based acceleration}

To go beyond the state-of-the-art, solid-state materials might offer new paths for beam manipulation and acceleration. For example, in the field of accelerator physics, the channelling properties of silicon crystal have successfully been used for collimation and extraction of relativistic proton beams \cite{crystalcollimation}. Could also solids provide an alternative medium for acceleration? Depending on their particular atomic configuration and electrical conduction nature, some solid-state micron and nano sized structures offer interesting properties to enhance  electric field components  or induce strong wakefields that could be useful for acceleration, as well as transverse particle guiding and radiation emission. Next we review some promising concepts that could revolutionise the future generation of accelerators and light sources. 

\subsection{Dielectric wakefield acceleration}

\begin{figure}[!htb]\centering
	\includegraphics[width=8cm]{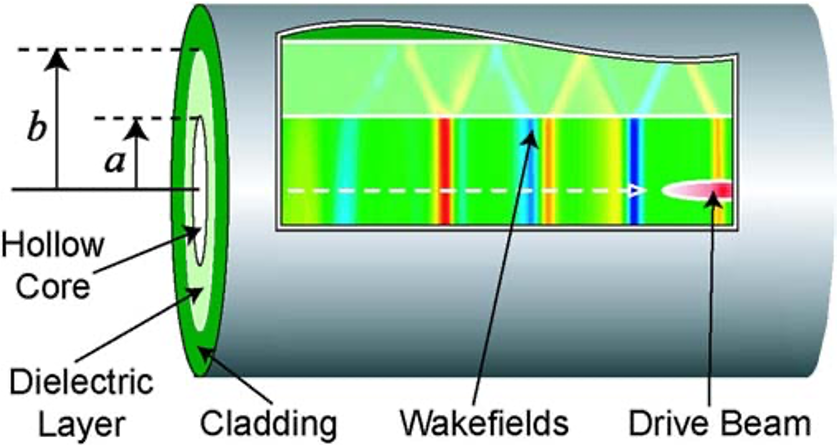}
	\caption{Conceptual drawing of DWA \cite{Thompson:2008}.}
\label{fig:DWA}
\end{figure}

Since the achievable peak field in the conventional RF metallic cavities is limited by surface breakdown, one obvious way to overcome this limit is through the use of materials with better breakdown properties, such as some dielectric materials (e.g. diamond, quartz, silica and ceramics).  Figure~\ref{fig:DWA} illustrates the concept of beam driven Dielectric Wakefield Acceleration (DWA) \cite{Thompson:2008}. 
In this case, the dielectric accelerator consists of a hollow channel covered by a layer of a dielectric material and a metallic cladding. The electric field from the driver polarizes the atoms of the
dielectric medium, which coherently generates high frequency electromagnetic radiation.

\begin{figure}[!htb]\centering
	\includegraphics[width=12cm]{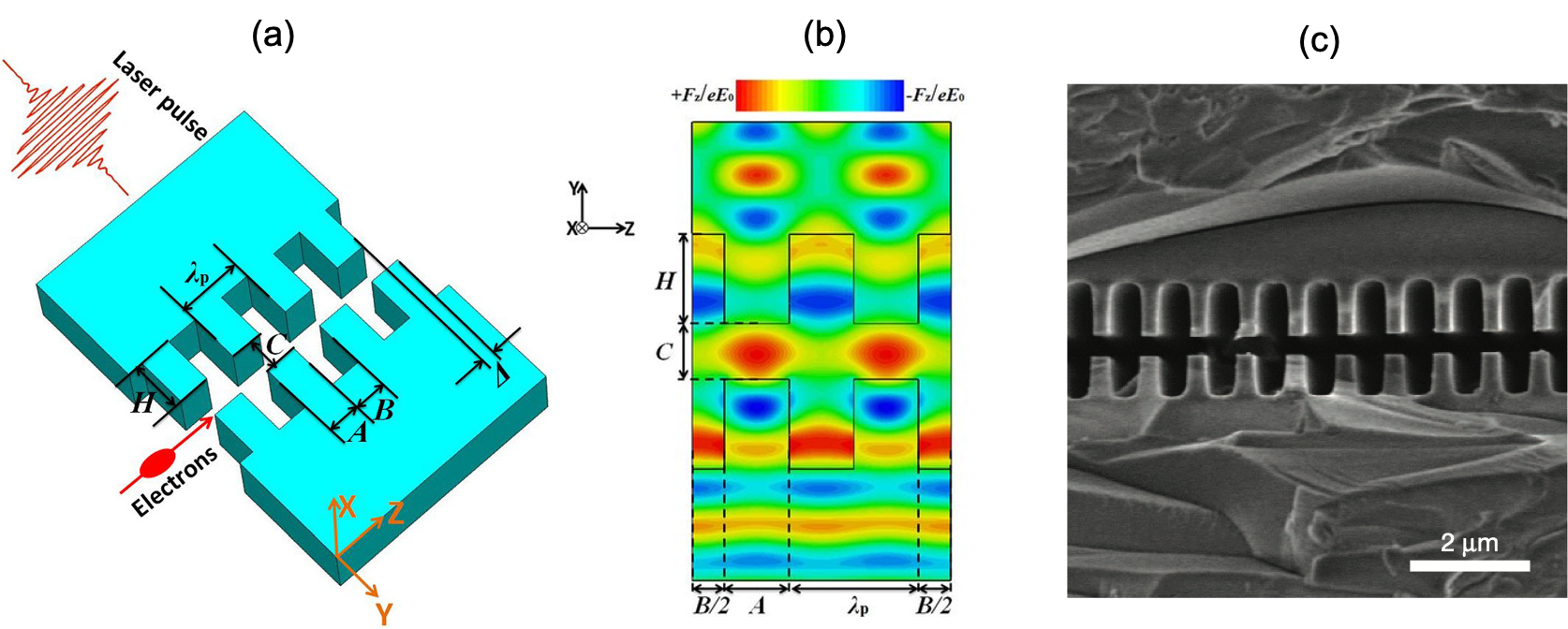}
	\caption{(a) Conceptual scheme of a dual-grating scheme of DLA. (b) Simulation of  the longitudinal accelerating force in a dual-grating structure illuminated by an input laser field along the y-axis \cite{Wei:2017}. (c) Scanning electron microscope image of the longitudinal cross-section of a DLA dual grating structure with 400 nm gap \cite{Peralta:2013}.}
\label{fig:DLA}
\end{figure}

Alternatively, lasers can be used as driving source. Dielectric Laser-driven acceleration (DLA) has been demonstrated for both non-relativistic \cite{Breuer:2013} and relativistic electrons \cite{Peralta:2013, OShea:2016} and has made enormous progress in recent years. DLA is the method behind the famous concept of "accelerator on a chip" \cite{Blau:2013, England:2021}. DLA and DWA are limited to a maximum gradient of approximately 10 GV/m in the THz frequency range. Figure~\ref{fig:DLA} shows a particular dual-grating DLA configuration with sub-micron aperture. 

\begin{figure}[!htb]\centering
	\includegraphics[width=10cm]{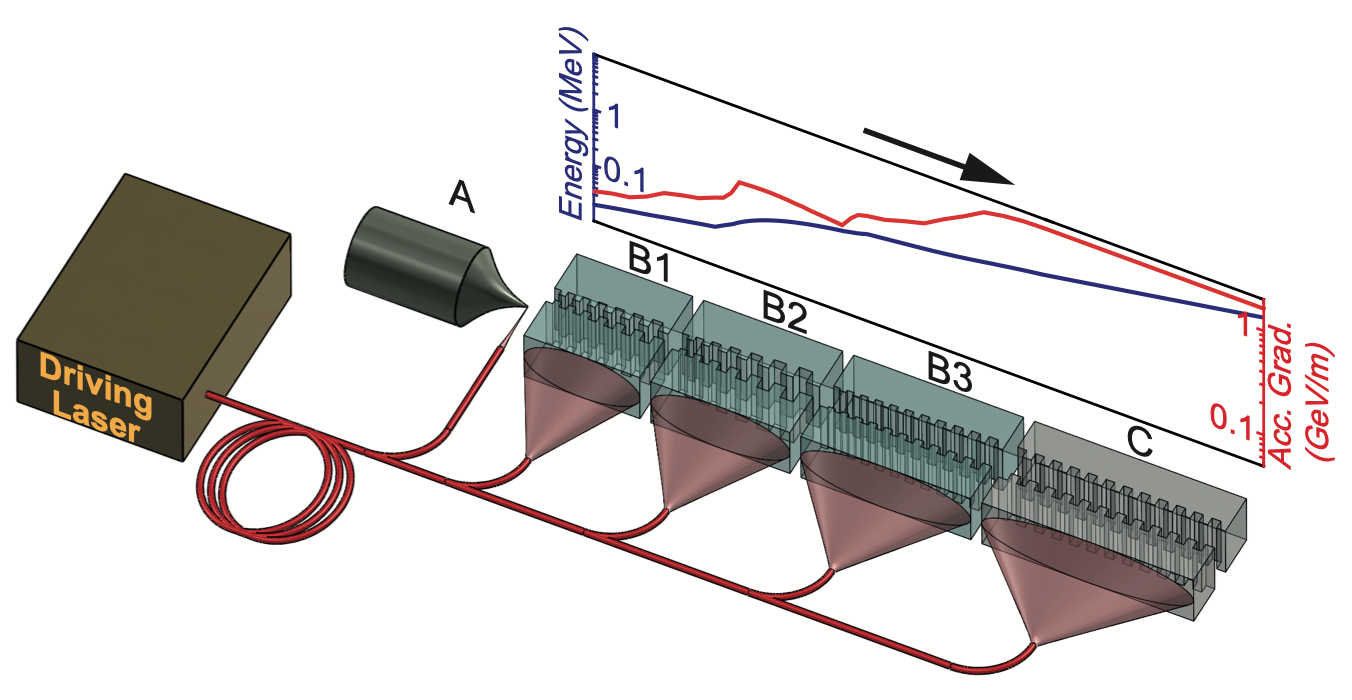}
	\caption{Conceptual design of a staged DLA collider, consisting of dielectric structures with a tapered grating period to guarantee synchronicity with the accelerating electrons \cite{Breuer:2013}.}
\label{fig:dielectriccollider}
\end{figure}

Usually the dielectric structures have sub-mm dimensions, and there have been conceptual proposals to build a high energy collider based on staged dielectric structures (Figure~\ref{fig:dielectriccollider}). This would allow "table-top" high energy accelerators. 

\subsection{Crystal channelling acceleration}

Semiconductor and metallic crystalline lattices have been proposed to generate a solid-state plasma medium to guide and accelerate positive particles, taking advantage of the channelling properties in crystals. High electron density in solids could be obtained from the conduction bands. Typical electron densities ($n_e$) in solid-state plasmas lie within the range of $10^{19}~\textrm{cm}^{-3} \le n_e \le 10^{24}~\textrm{cm}^{-3}$ \cite{Chen:1987,Ostling:1997}, i.e. between one and six orders of magnitude higher than the density in gaseous plasmas. Taking into account Eq.~(\ref{eq:longwake}), solid-state based plasmas might lead to accelerating gradients  $0.1~\textrm{TV/m} \lesssim E_z \lesssim 100~\textrm{TV/m}$. 

Solid-state wakefield acceleration using crystals was proposed in the 1980s and 1990s by T. Tajima and others \cite{Chen:1987, Tajima:1987, Chen:1997} as a technique to sustain TV/m acceleration gradients. In the original Tajima's concept \cite{Tajima:1987}, high energy ($\simeq 40$~keV) X-rays are injected into a crystalline lattice at the Bragg angle to cause Borrmann-Campbell effect \cite{Borrmann, Campbell}, yielding slow-wave accelerating fields. Then a witness beam of charged particles, e.g. muons, which is injected into the crystal with an optimal injection angle for channelling, can experience acceleration along the crystal axis. 

Alternatively, wakefields in crystals can be induced by means of the excitation of high frequency collective motion of conduction electrons through the crystalline lattice (see next section on plasmonic acceleration).  To reach accelerating gradients on the order of $\sim$TV/m, crystals must be excited by ultrashort X-ray laser pulses within a power range of TW-- PW, which makes the practical realisation of the concept very challenging. It has only recently become a realistic possibility since the invention of the so-called single-cycled optical laser compression technique by G. Mourou et al. \cite{Mourou:1985, Mourou:2014}. 

If natural crystals (e.g. silicon) are used for solid-state wakefield acceleration, the beam intensity acceptance is significantly limited by the angstrom-size channels. In addition, such small size channels increase the dechannelling rate and make the channels physically vulnerable to high energy interactions, thus increasing the damage probability by high power beams. 

Over the past decade there have been great advances in nanofabrication techniques \cite{PingLi, Cary, XChen} that could offer an excellent way to overcome many of the limitations of natural crystals. Metallic nanostructures and metamaterials \cite{Pizzi, Ling-Bao} may offer suitable ultra-dense plasma media for wakefield acceleration or charged particle beam manipulation, i.e. channelling, bending, wiggling, etc. This also includes the possibility of investigating new paths towards ultra-compact X-ray sources \cite{Pizzi}.  


\subsection{Plasmonic acceleration}

\begin{figure}[!htb]\centering
	\includegraphics[width=12cm]{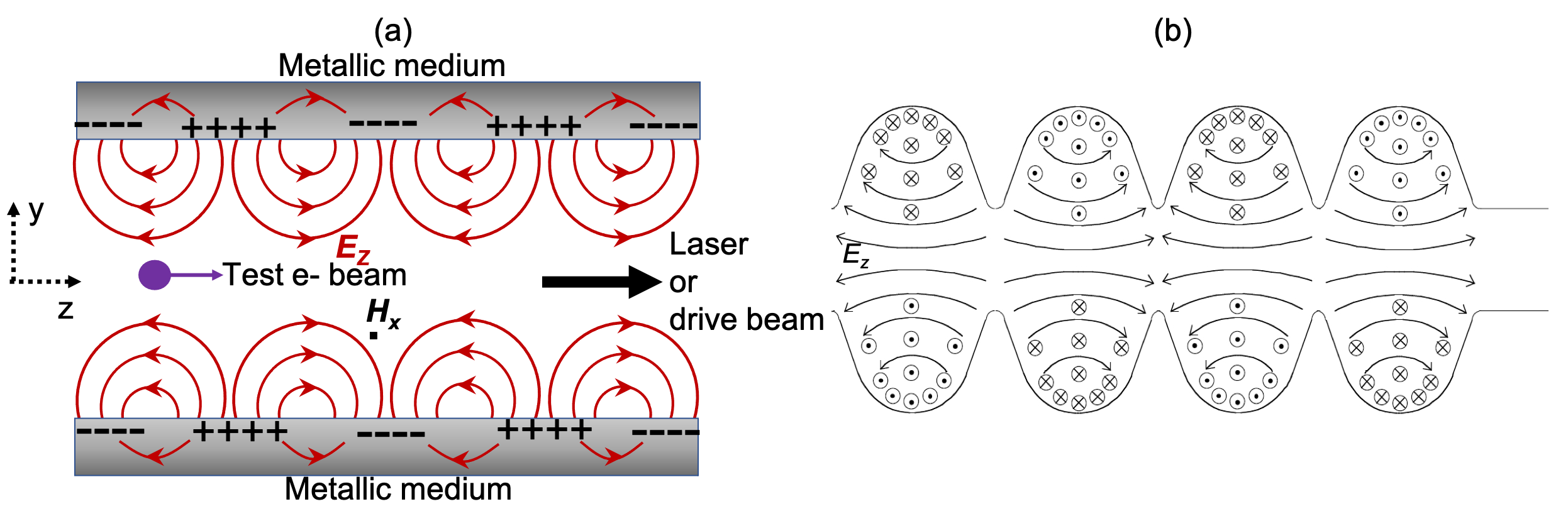}
	\caption{(a) Plasmonic acceleration concept. (b) Comparison with a RF cavity operating in $\pi$-mode. The drawings are not to scale. While the plasmonic structure has micrometric apertures and length on the order of 1 mm, a 9-cell RF cavity has usually apertures of tens of mm and length on the order of m.}
\label{fig:plasmon}
\end{figure}

Plasmonics can be defined as the study of the interaction between electromagnetic fields and the free electron Fermi gas in conducting solids. External electromagnetic fields can excite plasmons, i. e. collective oscillations of conduction electrons in metals \cite{Plasmons}. To some extent, this collective effect could be exploited to generate ultra-high acceleration gradients. Figure~\ref{fig:plasmon} depicts a scheme of excited plasmons in metallic surfaces. The oscillation of induced longitudinal electric field reminds that in a sequence of a multi-cell RF cavity operating at $\pi$-mode. 

\begin{figure}[!htb]
   \centering
   \includegraphics[width=14cm]{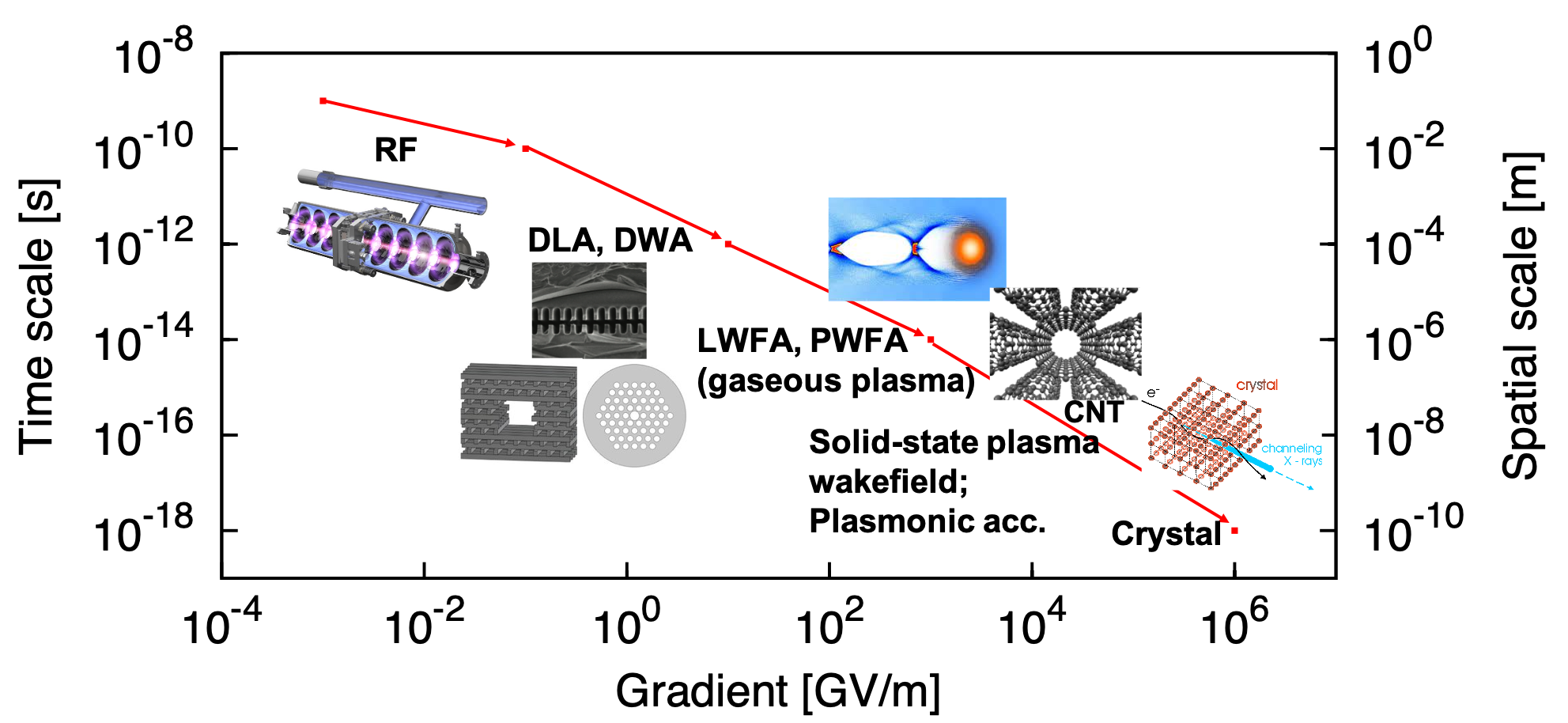}
   \caption{Schematic comparing the space and time scales of the longitudinal electric field components generated by different techniques for charged particle acceleration and their corresponding amplitude (acceleration gradient). The cases of nanostructure wakefields, plasmonic and crystal acceleration are based on theoretical and numerical predictions.}
   \label{fig:acctechniquescompara}
\end{figure}

The excitation of surface plasmonic modes \cite{surface_plasmonic1, surface_plasmonic2} can be driven either by charged particle beam \cite{Nejati:2009} or by laser \cite{Fedeli:2016}. To be effective, the driver dimensions should match the spatial ($\sim$ nm) and time (sub-femtoseconds) scales of the excited plasmonic oscillations. Wakefield driving sources working on these scales are now experimentally realizable. For instance, attosecond X-ray lasers are possible thanks to the pulse compression technique \cite{Mourou:2014}. In the case of beam-driven wakefields, the experimental facility FACET-II at SLAC \cite{FACETII} will allow the access to "quasi-solid" electron bunches with densities up to $10^{24}$ cm$^{-3}$ and submicron longitudinal size. A comparison of the range of spatial and time scales of the electric field oscillations and achievable acceleration gradient for standard and novel acceleration methods is shown in Figure~\ref{fig:acctechniquescompara}. In principle, solid-state wakefields and plasmonics acceleration with nanostructures are predicted to have the potential to generate higher acceleration gradients than DWA and DLA, LWFA and PWFA with gaseous plasma. Plasmonic acceleration with nanostructures could be considered an intermediate step between plasma wakefield acceleration and crystal channelling acceleration towards the PV/m gradient regime.

Due to their special thermo-mechanical and optoelectronic properties, materials based on carbon nanotubes arrays or graphene could offer an excellent medium to generate plasmonic wakefield acceleration. For instance, conduction electrons in CNTs could have densities $n_e \sim 10^{23}$~cm$^{-3}$. In principle, as proved in \cite{Rider:2012, Que:2002}, in the linear regime the plasmonic dynamics in CNT bundles can be described by classical plasma formulae. Therefore, from Eq.~(\ref{eq:longwake}) we can estimate a maximum longitudinal wakefield $E_z \sim 10$~TV/m. 

\begin{figure}[!htb]\centering
	\includegraphics[width=12cm]{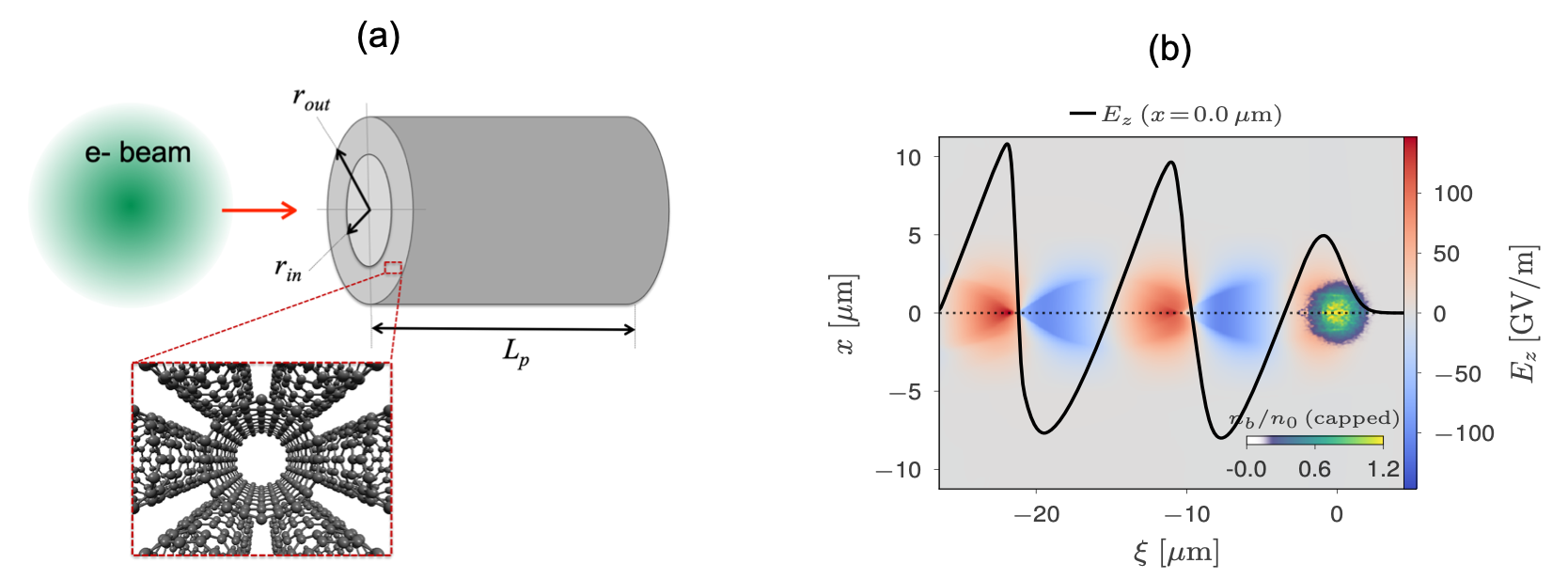}
	\caption{(a)Schematic model for beam-driven wakefield simulation using a hollow cylinder of solid-state plasma confined in a wall of thickness $r_{out} - r_{in}$ and length $L_p$. In this model the cylinder wall represents a solid made of CNT bundles. (b) An example of longitudinal wakefield as a function of the longitudinal coordinate $\xi = z-ct$, with $c$ the speed of light. This particular case has been computed using the PIC code FBPIC \cite{FBPIC}.}
\label{fig:CNTstructurewakefield}
\end{figure}

Particle-In-Cell (PIC) simulations of beam driven  wakefield acceleration in cylindrical metallic hollow structures (Figure~\ref{fig:CNTstructurewakefield}) with micrometric or nanometric apertures have shown the feasibility of obtaining gradients $\gtrsim 100$~GV/m, allowing envisioning  the possibility of an ultra-compact PeV linear collider \cite{Aakash:2020, Aakash:2021, Javier:2018}.  Similar gradient values have also been  predicted from PIC simulations of X-ray laser driven wakefield acceleration in nanotubes \cite{Zhang:2016}.


\subsection{Target normal sheath acceleration}

\begin{figure}[!htb]\centering
	\includegraphics[width=10cm]{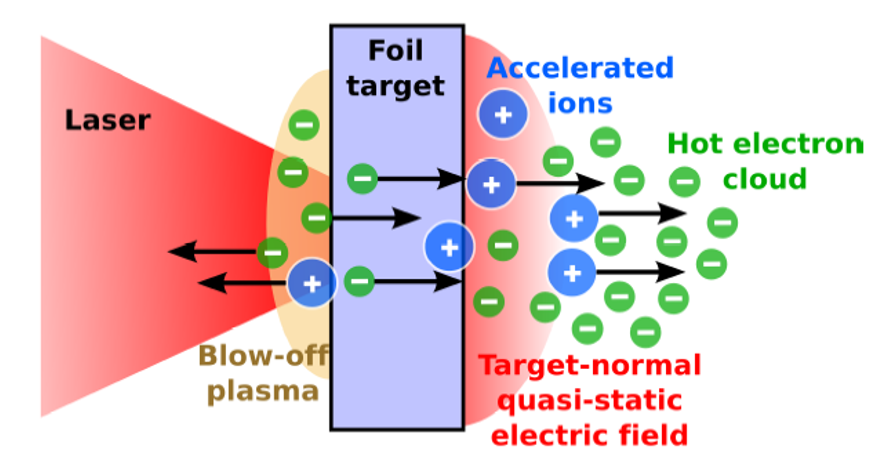}
	\caption{Target normal sheath acceleration mechanism.}
\label{fig:ablation}
\end{figure}

Solid-state plasmas can also be created by ablation of solids using high intensity near-infrared femtosecond lasers. In this case, unlike in the plasmon wakefield excitation, the individual ions are uncorrelated. For example, the so-called target normal sheath acceleration (TNSA) technique is based on the ablation of a metallic thin foil \cite{TNSA}. The mechanism is schematically represented in Figure~\ref{fig:ablation}.  A thin foil is irradiated by an intense laser pulse. The laser prepulse creates a pre-plasma on the target’s front side. The main pulse interacts with the plasma and accelerates MeV-energy electrons mainly in the forward direction. The electrons propagate through the target, where collisions with the background material can increase the divergence of the electron current. Then the electrons leave the rear side, resulting in a dense sheath. An electric field due to charge-separation is created. The field is of the order of the laser electric field ($\sim$TV/m),  which ionizes atoms at the surface. Eventually the ions are accelerated in this sheath field, pointing in the target normal direction.


\subsection{Solid-state plasma by electron field emission}

\begin{figure}[!htb]\centering
	\includegraphics[width=10cm]{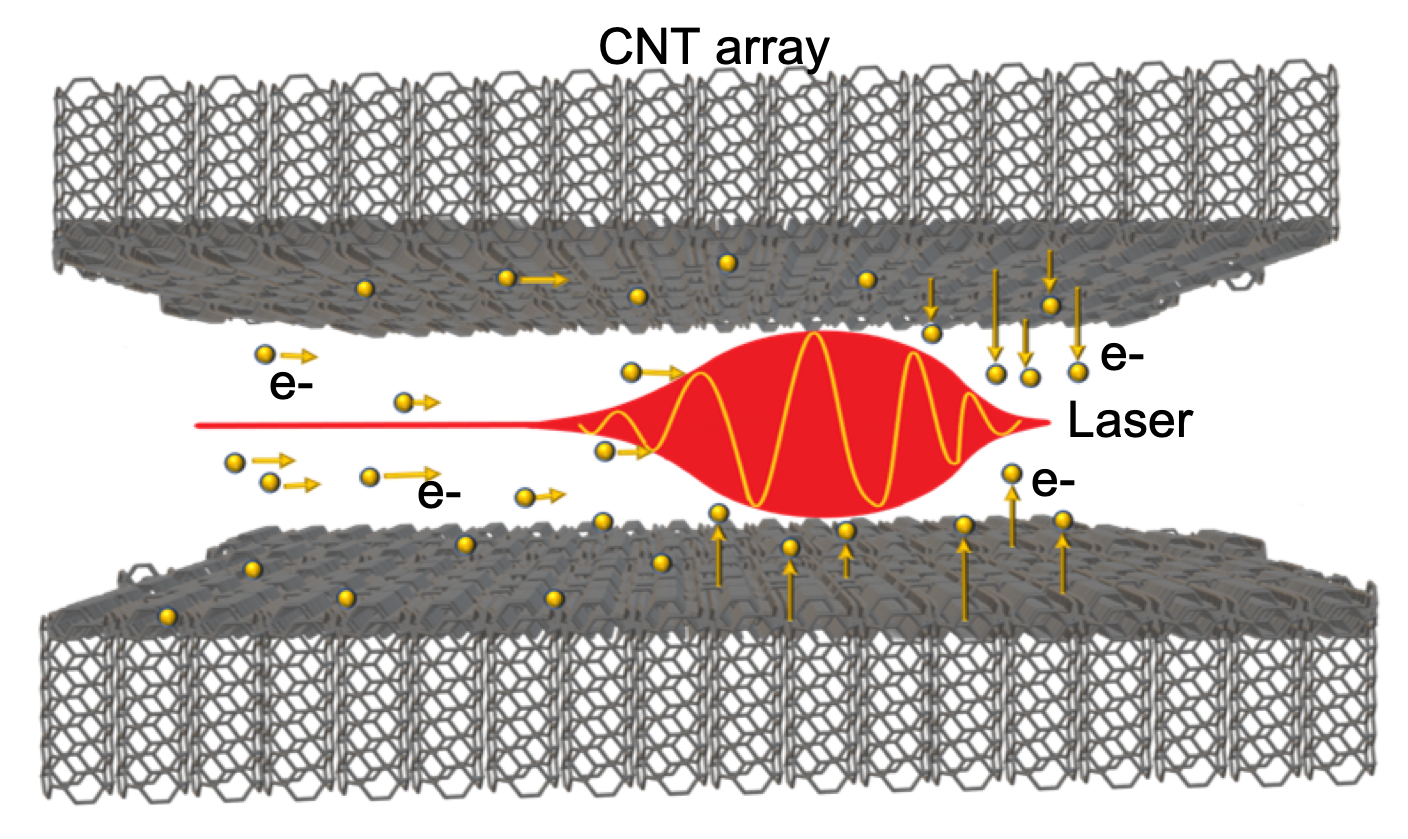}
	\caption{Laser wakefield acceleration concept with aligned CNT arrays. In this particular configuration CNTs are arranged perpendicularly to the laser propagation \cite{Nick:2021}.}
\label{fig:CNTemission}
\end{figure}

Electron field emission from nanostructures could offer another interesting way to generate solid-state plasma. A possible configuration based on two layers of aligned CNT arrays and a gap in between them  is shown in Figure~\ref{fig:CNTemission}. Taking into account the low work function of CNTs and their emission properties, a first intense laser pulse, traveling through the gap between the CNT sheets, could be used to induce electron field emission from CNT arrays. Then extracted electrons in the gap are accelerated by the ponderomotive force of a second laser pulse.  For more details see Ref.~\cite{Nick:2021}.

\section{Outlook and prospects}

Currently particle accelerators play an important role in many areas, including fundamental science, medicine, industry and society in general. During the last 100 years, the research in HEP has been the main driving force behind the development and progress of the accelerator technology. The LHC is currently the world's most advanced and powerful accelerator, and the only multi-TeV collider in operation. It consists of a 27-km ring of superconducting magnets with a number of accelerating structures based on RF electromagnetic fields to accelerate particles. The next generation of RF-driven high energy accelerators will pose serious challenges in terms of infrastructure size, technology and cost.

The current state-of-the-art of the RF techniques is limited to gradients on the order of 100 MV/m. Hence, larger and more expensive accelerator facilities are necessary in order to go beyond the LHC capabilities. Therefore, it is important to explore alternative advanced acceleration techniques to overcome the limitations of conventional technologies, thus progressing towards more compact, sustainable and economical particle accelerators. For many years, the development of inexpensive and compact "table-top" high energy particle accelerators and light sources has been the ultimate dream of the accelerator physics community.

Some of the most promising novel acceleration techniques have briefly been reviewed in this article. For instance,  PWFA and LWFA using gaseous plasma have experimentally demonstrated to be able to achieve $\sim 100$~GV/m acceleration \cite{Gordon:1998, Leemans:2014, Litos:2015, Blue:2003, Corde:2015, Doche:2017}. Another path towards TV/m acceleration is the use of special  solid-state media, such as micrometric and millimetric grating dielectric structures, reaching  $\sim 1$--10~GV/m acceleration \cite{Peralta:2013, OShea:2016}. 

Crystals and, more recently, nanostructures have also attracted attention to create a solid-state plasma medium with electron densities 1--6 orders of magnitude higher than those in gaseous plasmas. For example, recent theoretical and numerical studies have shown the feasibility of obtaining tens of TV/m fields in nanomaterials by excitation of non-linear plasmonic modes  driven by a high density sub-micron particle bunch \cite{Aakash:2020, Aakash:2021}. To obtain multi-TV/m fields other techniques rely on the ablation of solids \cite{TNSA} or the induced electron field emission from nanomaterials \cite{Nick:2021}.

It is necessary to remark that the excitation of solid-state plasma wakefields requires the use of driving sources (lasers or charged particle bunches) in the same time and spatial scales as the wakefield oscillations in the plasma. In principle, wakefield driving sources working  on solid-state plasma scales are now experimentally realizable. Attosecond X-ray lasers are possible thanks to the pulse compression \cite{Mourou:1985, Mourou:2014}. In the case of beam-driven wakefields, the experimental facility FACET-II at SLAC \cite{FACETII} could deliver "quasi-solid" electron beams in the near future. Therefore, an experimental proof-of-concept of solid-state plasma wakefield acceleration could possibly become a reality within the next 10 years. 

It is also worth mentioning that nanostructures and metamaterials, due to their special and flexible optoelectronic properties, could also provide an excellent medium to create compact X-ray sources, eg. X-ray light sources based on CNT arrays \cite{Shin:2017}.

In conclusion, the different advanced acceleration techniques described in this article have the potential to transform the current paradigm in accelerator physics. They offer novel pathways to access the multi-GV/m  and multi-TV/m field regimes. This will open new horizons to the physics of extreme fields, particularly in collider physics, light sources, and in many other areas of applied sciences, medicine and industry.


\bibliography{references}  


\end{document}